\pdfoutput=1
\documentclass[
reprint,
superscriptaddress,
amsmath,amssymb,
aps, prl,
floatfix,
]{revtex4-2}
\usepackage{graphicx}
\usepackage{dcolumn}
\usepackage{bm}
\usepackage{hyperref}
\usepackage[mathlines]{lineno}
\usepackage{here}
\usepackage{siunitx}
\usepackage{mhchem}
\usepackage{tabularx}
\usepackage{booktabs}
\usepackage{adjustbox}
\begin{document}
	
	\preprint{APS/123-QED}
	
	\title{ Exploring Li-ion Transport Properties of Li$_3$TiCl$_6$: A Machine Learning Molecular Dynamics Study} 
	
	\author{Selva Chandrasekaran Selvaraj}
	\affiliation{Department of Chemical Engineering, University of Illinois Chicago, Chicago, IL 60608, United States}
	\affiliation{Materials Science Division, Argonne National Laboratory, Lemont, IL 60439, United States}   
	\author{Volodymyr Koverga}
	\affiliation{Department of Chemical Engineering, University of Illinois Chicago, Chicago, IL 60608, United States}
	\affiliation{Materials Science Division, Argonne National Laboratory, Lemont, IL 60439, United States}   
	\author{Anh T. Ngo}
	\email{anhngo@uic.edu}
	\affiliation{Department of Chemical Engineering, University of Illinois Chicago, Chicago, IL 60608, United States}
	\affiliation{Materials Science Division, Argonne National Laboratory, Lemont, IL 60439, United States}

	\date{\today}

	\begin{abstract}
			We performed large-scale molecular dynamics simulations based on a machine-learning force field (MLFF) to investigate the Li-ion transport mechanism in cation-disordered Li$_3$TiCl$_6$ cathode at six different temperatures, ranging from 25$^\mathrm{o}$C to 100$^\mathrm{o}$C. In this work, deep neural network method and data generated by $ab-initio$ molecular dynamics (AIMD) simulations were deployed to build a high-fidelity MLFF. Radial distribution functions, Li-ion mean square displacements (MSD), diffusion coefficients, ionic conductivity, activation energy, and crystallographic direction-dependent migration barriers were calculated and compared with corresponding AIMD and experimental data to benchmark the accuracy of the MLFF. From MSD analysis, we captured both the self and distinct parts of Li-ion dynamics. The latter reveals that the Li-ions are involved in anti-correlation motion that was rarely reported for solid-state materials. Similarly, the self and distinct parts of Li-ion dynamics were used to determine Haven's ratio to describe the Li-ion transport mechanism in Li$_3$TiCl$_6$. Obtained trajectory from molecular dynamics infers that the Li-ion transportation is mainly through interstitial hopping which was confirmed by intra- and inter-layer Li-ion displacement with respect to simulation time. Ionic conductivity (1.06 mS/cm) and activation energy (0.29eV) calculated by our simulation are highly comparable with that of experimental values. Overall, the combination of machine-learning methods and AIMD simulations explains the intricate electrochemical properties of the Li$_3$TiCl$_6$ cathode with remarkably reduced computational time. Thus, our work strongly suggests that the deep neural network-based MLFF could be a promising method for large-scale complex materials.
	\end{abstract}
	
	% (optional)
	
	\maketitle 

	\section{Introduction}
	
	Addressing the increasing energy demands in the face of climate change concerns requires a sustainable zero-carbon energy future. Rechargeable batteries that are capable of converting electrical energy to chemical energy and vice versa, are pivotal for energy storage in this context. While lithium-ion batteries have proven successful for portable devices, all-solid-state batteries (ASSLBs) present a promising solution for the next generation. ASSLBs offer enhanced safety and a longer lifespan compared to traditional lithium-ion electric vehicle batteries and are aligning with goals for achieving zero-carbon emissions.
		\begin{figure*}[htbp]
		\centering
		\includegraphics[width=0.9\textwidth]{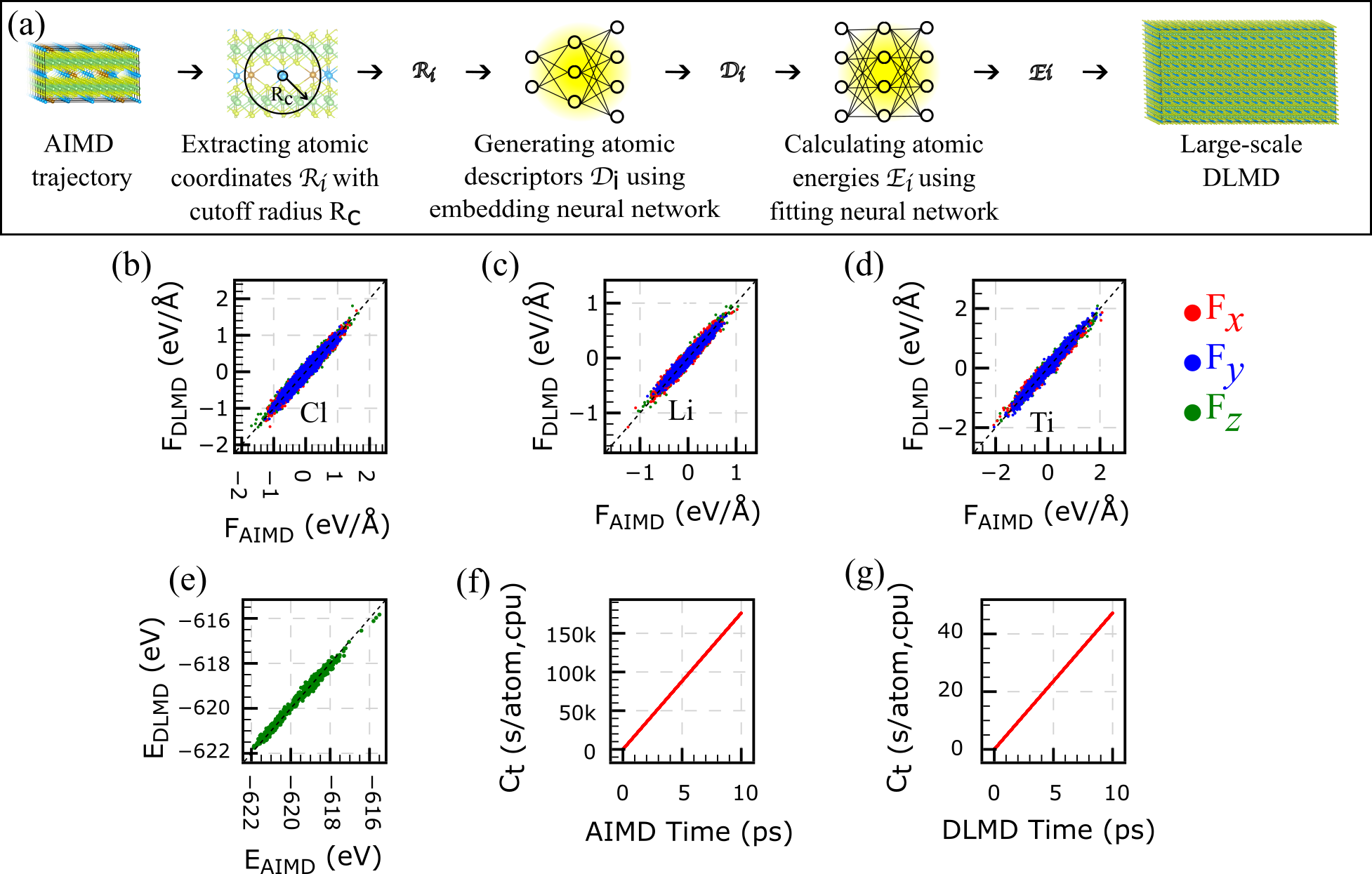}
		\vspace{2mm}
		\caption{(a) Schematic diagram of the protocol of DLMD development (see Section 2 for the details).  (b-d) DLMD predicted forces corresponding to AIMD forces for Cl (b) , Li (c) and Ti (d) elements, as well as the respective forces along $x$, $y$, and $z$ directions.  (e) DLMD predicted energies corresponding to AIMD energies. (e-f) Simulation time (C$_t$) of AIMD(e) and DLMD(f) approaches, respectively. }
		\label{fig:Figure1}
	\end{figure*}	
	
	In ASSBs, the cathode is a crucial component and it is playing a vital role in determining their overall performance. Particularly, energy storage capacity, voltage, cycle life, safety, cost, and environmental impact, are intricately linked to the cathode material. Among the widely used sulfide- and oxide-based cathode materials, latter take a leading position in energy storage manufacturing. Indeed, such commercial attention is also facilitated by the use of transition metals, including single-  Li$M$O$_2$, multi-  LiNi$_{1-x-y}$Mn$_x$Co$_y$O$_2$, and polyanionic-TM-based Li$M$PO$_4$ ($M$=3$d$-block elements of 3 group; x =y=0.1 to 0.33),\cite{cd1, cd2, cd3, cd4, cd5, cd6, cd7, cd8, cd9, cd10} which essentially improve the performance of ASSLB devices. Since the projected annual production of Li-ion batteries is expected to reach several terawatt hours, demand for Fe, Co, and Ni based cathode materials increases rapidly\cite{tm}. However, Co and Ni prove to be expensive. Consequently, ongoing efforts include both simulations and experimental studies, aiming to identify more cost-efficient alternatives for cathodes\cite{dft2}.
	
	Among various studies on cathode materials, a recent research has shown that Li$_3$TiCl$_6$ is both cost-effective and outperforms previous benchmarks for ASSLBs\cite{wang2023li3ticl6}. However, the experimental investigation \cite{wang2023li3ticl6} of Li$_3$TiCl$_6$ falls short of providing a comprehensive understanding of the underlying physics and chemistry governing the Li-ion transport mechanism, which is a key factor to determine the performance entire battery. Thus, we turn to atomic simulation techniques to study the underlying Li-ion transport mechanism in Li$_3$TiCl$_6$.
	
	The conventional atomic simulation based on density functional theory (DFT) is well known for predicting structural, electrochemical, and Li-ion transport properties \cite{dft1, dft2} with few hundred atoms. Nevertheless, it requires extensive computational resources and is helpless for large-scale demand posed by Li-ion intercalation-driven electrochemical studies. To facilitate this challenge, various machine learning (ML) methods are utilized in conjunction with molecular dynamics simulations\cite{Schleder_2019, Schleder_2020}. Among the variety of ML techniques, e.g., artificial neural networks,\cite{batzner20223, Wen_2022} kernel-based methods,\cite{Rupp}, Gaussian approximation potentials,\cite{Bart} and atomic cluster expansion,\cite{Drautz} Deep Learning Potential (DLP) stands out as versatile tool, capable of producing accurate potential models on the basis of quantum-chemical calculations \cite{Peng_2023, Zhang}. % At the same time, the effectiveness and precision of ML potentials hinge on the quality and size of the training dataset .
	
	Therefore, our approach in this work involves the integration of AIMD, machine learning methods based on deep neural networks, and classical molecular dynamics. This integrated approach is collectively referred to as deep learning molecular dynamics (DLMD) simulations \cite{Peng_2023, Zhang} to investigate the structural and Li-ion transport properties of the Li$_3$TiCl$_6$ cathode. With this, we organize the manuscript as follows: Section II describes detailed simulation techniques, as illustrated in FIG.\ref{fig:Figure1}. Section III is composed of results and discussions covering structural parameters, RDF, MSD, self and correlated motion of Li-ion displacement, diffusion coefficient, Li-ion transport mechanism, ionic conductivity, and activation energy of Li$_3$TiCl$_6$. The calculated values of ionic conductivity and activation energy are in good agreement with those of experimental values \cite{wang2023li3ticl6}. Finally, we conclude our results and discussions in Section IV.
	
	\section{Simulation Methodologies}
	\subsection{AIMD simulation}
	DFT-based calculations were employed to optimize the crystal structure of  Li$_3$TiCl$_6$ using the Vienna $Ab-initio$ Simulation Package (VASP)\cite{vasp1, vasp2}. The initial structure of  Li$_3$TiCl$_6$ in monoclinic cell with $C2/m$ space group for VASP was adopted from experimental results\cite{wang2023li3ticl6}. The projector augmented wave (PAW) formalism described the valence electrons of Li, Cl, and Ti atoms using plane wave-based wave functions was employed\cite{paw}. The structure optimization, involving the minimization of ground state energy, utilized the generalized gradient approximation of Perdew and Wang method with different $U$-parameters\cite{gga, gga1}. A set of $U-J$ values, 0 and 4eV were chosen to taking into account of strong correlation effect of Ti-3$d$ electrons\cite{u_para, u_para1}. A kinetic cutoff energy of 450 eV was set to enhance calculation accuracy. The Brillouin zone of the supercell with 160 and 320 atoms was sampled with 4$\times$2$\times$4 and 4$\times$2$\times$2 $k$-meshes for ionic optimization, respectively. Ionic and electronic optimizations were alternately performed until the forces on each ion reached less than ±10 meV/Å. 
	
	The designed simulation cells were optimized and subjected to AIMD simulations. The time step for the AIMD simulations was set to 1 fs, and the simulations were continued for a duration considered separately within the canonical NVT ensemble to generate the dataset for the DLP  model.
	\begin{table}[h]
		\centering
		\caption{ The calculated loss function parameter expressed by mean absolute error (MAE) and  root mean square error (RMSE) of deep learning potential model}
		\begin{tabularx}{\linewidth}{XcXcXcX}
			\toprule
			\textbf{Metric} & \textbf{Value} & \textbf{Unit} \\
			\midrule
			Energy MAE    & $6.723 \times 10^{-4}$ & eV/atom \\
			Energy RMSE & $8.389 \times 10^{-4}$ & eV/atom \\
			\midrule
			Force MAE   & $5.959 \times 10^{-3}$ & eV/\AA \\
			Force RMSE & $7.827 \times 10^{-3}$ & eV/\AA \\
			\bottomrule
		\end{tabularx}
	\end{table}

	\begin{table*}[htbp]
		\centering
		\caption{Calculated lattice parameters of monoclinic Li$_3$TiCl$_6$ with different Ti occupancy at 4$h$ and 2$a$ sites.}
		\setlength{\tabcolsep}{8pt} 
		\begin{tabular}{ccccccccc}
			\toprule
			\multicolumn{1}{c}{} & 
			\multicolumn{2}{c}{\shortstack{0.875Ti@2$a$\\0.063Ti@4$g$}} &
			\multicolumn{1}{c}{\shortstack{0.851Ti@2$a$\\0.075Ti@4$g$}} &
			\multicolumn{2}{c}{\shortstack{0.844Ti@2$a$\\0.078Ti@4$g$}} &
			\multicolumn{1}{c}{\shortstack{0.754Ti@2$a$\\0.123Ti@4$g$}} &
			\multicolumn{2}{c}{\shortstack{0.750Ti@2$a$\\0.125Ti@4$g$}} \\
			\cmidrule(lr){2-3} 
			\cmidrule(lr){4-4} 
			\cmidrule(lr){5-6} 
			\cmidrule(lr){7-7} 
			\cmidrule(lr){8-9} 
			\multicolumn{1}{c}{} &
			\multicolumn{1}{c}{\centering \shortstack{U$_{eff}$\\ =0.0}} & 
			\multicolumn{1}{c}{\centering \shortstack{U$_{eff}$\\ =4.0}} &  
			\multicolumn{1}{c}{\centering Expt \cite{wang2023li3ticl6}} &  
			\multicolumn{1}{c}{\centering \shortstack{U$_{eff}$\\ =0.0}} & 
			\multicolumn{1}{c}{\centering \shortstack{U$_{eff}$\\ =4.0}} & 
			\multicolumn{1}{c}{\centering Expt \cite{wang2023li3ticl6}} &
			\multicolumn{1}{c}{\centering \shortstack{U$_{eff}$\\ =0.0}} & 
			\multicolumn{1}{c}{\centering \shortstack{U$_{eff}$\\ =4.0}} \\ 
			\cmidrule{1-9} 
			Lattice constant, $a$(Å) & 6.434 & 6.401 & 6.350 & 6.415 & 6.396 & 6.350 & 6.451 & 6.398 \\
			Lattice constant, $b$(Å)  & 11.05 & 10.91 & 10.88 & 10.96 & 11.93 & 10.89 & 11.02 & 10.95  \\
			Lattice constant, $c$(Å)   & 6.351 & 6.340 & 6.337 & 6.383 & 6.348 & 6.353 & 6.413 & 6.381 \\
			Angle $\beta(^o)$ & 110.7 & 110.6 & 110.2 & 110.6 & 110.4 & 110.1 & 110.9 & 110.8  \\
			\bottomrule
		\end{tabular}
		%\end{tabular}
		\label{tab:example}
	\end{table*}
	\subsection{Deep Learning Potential}
	The DLP was developed using the deep neural network method implemented in DeePMD-kit (v2.2.7) \cite{zeng2023deepmd}. The deep neural network algorithm in DeePMD is designed using the TensorFlow Python library  \cite{tensorflow}. The deep potential-smooth edition (DeepPot-SE), an end-to-end machine learning-based potential energy surface (PES) model, was employed with a cutoff radius of 7 Å to include more neighbour atoms in the featurization process. Indeed, it efficiently represents the PES of a wide variety of systems with the accuracy of $ab-initio$ models \cite{DeepPot-SE}.
	
	In the process of constructing DLP, local coordinate matrix, $\boldsymbol{\mathcal{R}}$ and local atomic environment matrix,  $\left\{\boldsymbol{\mathcal{R}}_{ij}\right\}_{i=1}^{N} $ are represented as shown in Eq. (1) and (2).	
	\begin{equation}{\label{Eq:Ri}}
		\boldsymbol{\mathcal{R}}  = \left\{ \mathbf{r}_1^T, \ldots, \mathbf{r}_i^T, \ldots, \mathbf{r}_N^T \right\}^T
	\end{equation}
	where $\mathbf{r}_i = (x_i, y_i, z_i)$ which contains 3 Cartesian coordinates of atom and  $N$ is total number of atoms. And $\boldsymbol{\mathcal{R}}$ can be transformed into local environment matrices as
		\begin{equation}
		\begin{split}
			\left\{ \boldsymbol{\mathcal{R}}_{ij} \right\}_{i=1}^{N} &= \left\{\boldsymbol{r}_{i1}^{T}, \cdots, \boldsymbol{r}_{i2}^{T}, \cdots, \right. \\
			&\left. \boldsymbol{r}_{ij}^{T} \,|\, j \in N_{R_c}(i, R_c=7.0)\right\}^T 
		\end{split}
	\end{equation}
	
	where $j$ and $N_{R_c}(i)$ are indexes and number of neighbors of $i^{th}$ atom within the cut-off radius $r_c = 7.0$Å, and $\boldsymbol{r}_{j i} \equiv \boldsymbol{r}_{j}-\boldsymbol{r}_{i}$
	is defined as the relative coordinate. 
	
	An embedding neural network with three layers, each containing 32, 64, and 128 neurons, was used to convert the local atomic coordinates ($\boldsymbol{\mathcal{R}} \in \mathbb{R}^{N \times 3}$) into atomic descriptors $ \mathcal{D}i = \mathcal{D}i \left( \left\{ \boldsymbol{\mathcal{R}}_{ij} \right\}_{i=1}^{N} \right) $ that preserves the structural symmetries of the system \cite{zeng2023deepmd, WANG2018178}. Descriptor values were input into a fitting neural network with three layers, each comprising 256 neurons, that maps descriptors to atomic energies $E_i$ \cite{zeng2023deepmd, WANG2018178}, and thus the total energy and force of each atom in the system are calculated by Eq. (3) and Eq. (4).
	\begin{equation}{\label{Eq:energy}}
		E = \sum_{i=0}^{i=N} E_i = \sum_{i=0}^{i=N} E\left(\mathcal{D}_{i}\right)
	\end{equation}
	\begin{equation}{\label{Eq:force}}
		F_{i,\alpha} = -\frac{\partial E}{\partial r_{i, \alpha}}
	\end{equation}
	
	The hyperbolic tangent (\(\tanh\)) activation function was applied in the neural network, introducing non-linearity to effectively train the intricate atomic descriptor data \(\mathcal{D}_i\). The training process consisted of \(10^4\) steps, utilizing mean squared error for both energies and forces as the loss function at each training step. The optimization was facilitated by the Adam optimizer, initiating with a learning rate of \(10^{-3}\) and concluding at \(10^{-8}\), with a decay parameter set to 5000. With these specified parameters, a dataset comprising 1.3 million training samples and 10000 test samples from diverse trajectories were employed to construct the DLP model. The calculated loss function parameters such as mean absolute error and root mean square error during validation of developed DLP model is tabulated in Table-1. Accuracy of the predicted data of force and energy is more than 99 \% and data points of predicted versus trained are shown in FIG.\ref{fig:Figure1}b to FIG.\ref{fig:Figure1}e. 
	
	\subsection{Deep learning molecular dynamic simulation}
	Similarly to AIMD conditions, The simulations of DLMD was conducted in the NVT ensemble using the LAMMPS simulation package \cite{lammps} coupled with the DeepMD plugin \cite{WANG2018178}. The simulation cell was designed to accommodate 20,000 atoms. Simulations were carried out at six different temperatures—298, 313, 328, 333, 358, and 373 K—controlled by the Nose-Hoover thermostat. 
	
	Prior to NVT simulation, the energy minimization was conducted using  conjugate gradient algorithm. The temperature damping parameter was set to 100 fs, and a uniform integration time-step of 1 fs was employed for all simulations, extending over a total duration of 5.5 ns. Computing time for AIMD and DLMD simulation for 10 ps per atom on one computing core was calculated and plotted in FIG.\ref{fig:Figure1}g and FIG.\ref{fig:Figure1}g, respectively and it reveals that DLMD is 3730 times faster than AIMD simulation. The unit-cell of AIMD simulations were visualized using VESTA software\cite{vesta} and the trajectories of AIMD and DLMD simulations were visualized using OVITO software.\cite{Ovito} 
	\begin{figure}[htbp]
		\centering
		\includegraphics[width=0.45\textwidth]{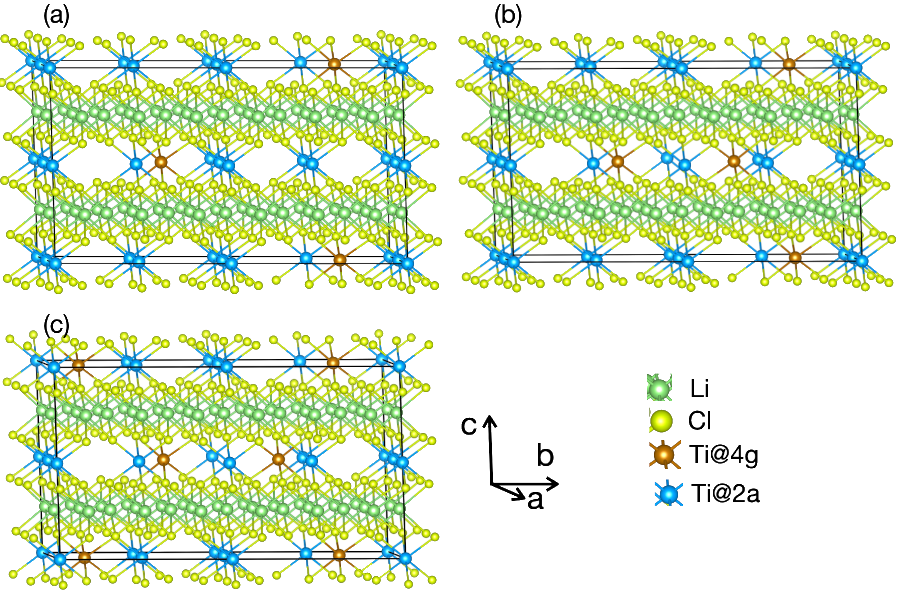}
		\vspace{2mm}
		\caption{Ball and stick model of Li$_3$TiCl$_6$ structure with three different Ti occupancies at 2$a$ and 4$g$ sites. }
		\label{fig:Figure2}
	\end{figure}
	\section{Results and Discussion}
	\subsection{Radial distribution}
	The monoclinic simulation cells, comprising 160 and 320 atoms, were employed to create three distinct Ti occupancies at the 2$a$ and 4$g$ sites, aiming to reproduce the experimentally reported structure of Li$_3$TiCl$_6$ \cite{wang2023li3ticl6} (FIG.\ref{fig:Figure2}). The three different occupancies, along with the corresponding experimental Ti occupancy, are summarized in Table-2. 
	
	Subsequently, the designed cells underwent optimization, and their structural parameters closely matched experimental values\cite{wang2023li3ticl6} for DFT+U calculation with U=4.0 eV. The slight variations, when compared to their experimental values, may be attributed to the small change in the Ti occupancy within the designed simulation cells as well as the overestimation of GGA functional that used in the DFT simulation.  Based on the structural optimization results, we selected the Li$_3$TiCl$_6$ structure with Ti-sites at 0.754 on 2$a$ and 0.123 on 4$g$, which represents experimentally annealed structure at 300$^\mathrm{o}$C and exhibited a maximum Li-ion conductivity of 1.04 mS/cm at room temperature\cite{wang2023li3ticl6} .

	\begin{figure}[htbp]
		\centering
		\includegraphics[width=0.37\textwidth]{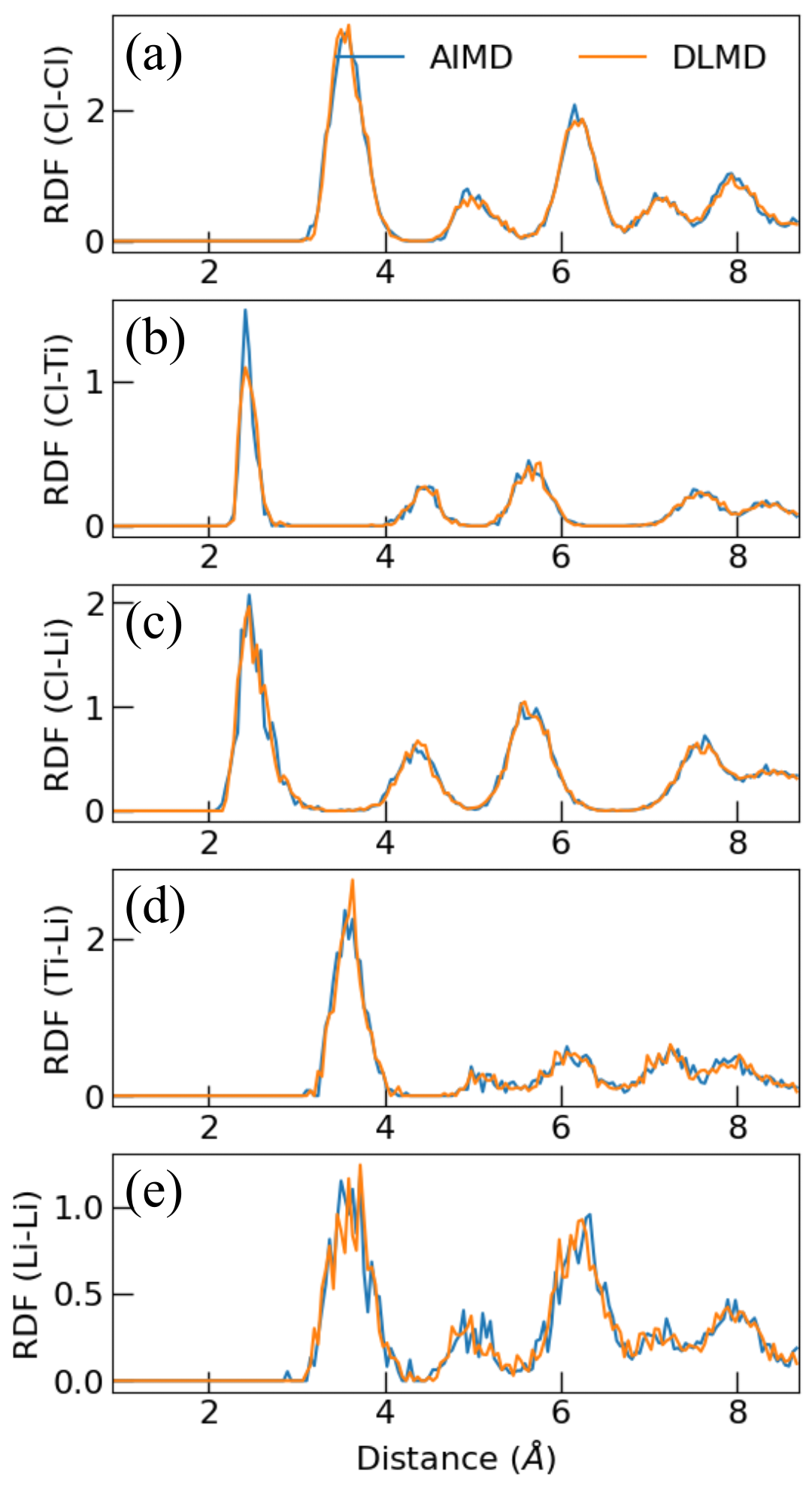}
		\caption{Comparison of radial distribution function of Li$_3$TiCl$_6$ analyzed based on AIMD with U=4.0 eV and DLMD trajectories at 298K.}
		\label{fig:Figure3}
	\end{figure}
	
	Atomic trajectories produced by AIMD and DLMD at 298 K were compared in the framework of radial distribution function (RDF), calculated by the equation given below:
	\begin{equation}{\label{Eq:rdf}}
		g(r) = \frac{1}{N} \sum_{i=1}^{N} \sum_{j \neq i} \frac{1}{4\pi r_{ij}^2 \Delta r} \delta(r - r_{ij})
	\end{equation}
	
	where, $N$, $r_{ij}$, and $\Delta r$ represent the total number of atoms in the radius $r$, the distance between atoms $i$ and $j$, and the width of each bin, respectively. 
	
	Generally, RDFs between pairs of atoms in the same material, estimated using AIMD and DLMD, may provide valuable confirmation of reliability, accuracy, and consistency in capturing the structural features of the system under investigation. Thus, results of the RDF analyzed using two theoretical approximations for Cl-Ti, Cl-Li, Ti-Li, Li-Li, and Cl-Cl pairs of atoms are presented in FIG.\ref{fig:Figure3}. As can be clearly seen, all the peak positions, shapes, and heights of RDF obtained by means of DLMD exhibit good agreement with AIMD results, even at larger distance, confirming the ability of DLMD to predict other static properties with high-level accuracy.
	
	Considering the RDF results more carefully, one can observe multiple peaks along the large separation distances which indicates a relatively high degree of atomic structuring in Li$_3$TiCl$_6$. For Cl-Ti and Cl-Li pairs, the first peak position is nearly the same at a distance of 2.53 Å, while other pair combinations are situated at larger distances of 3.5 Å. At the same time, the height of the Li-Cl peak is greater than that of Ti-Cl, which reveals that a relatively higher probability of Li-Cl interactions compared to Ti-Cl. This is evident when comparing the peak heights of Li-Li, Ti-Li, and Cl-Cl pairs.
	
	\begin{figure}[htbp]
		\centering
		\includegraphics[width=0.5\textwidth]{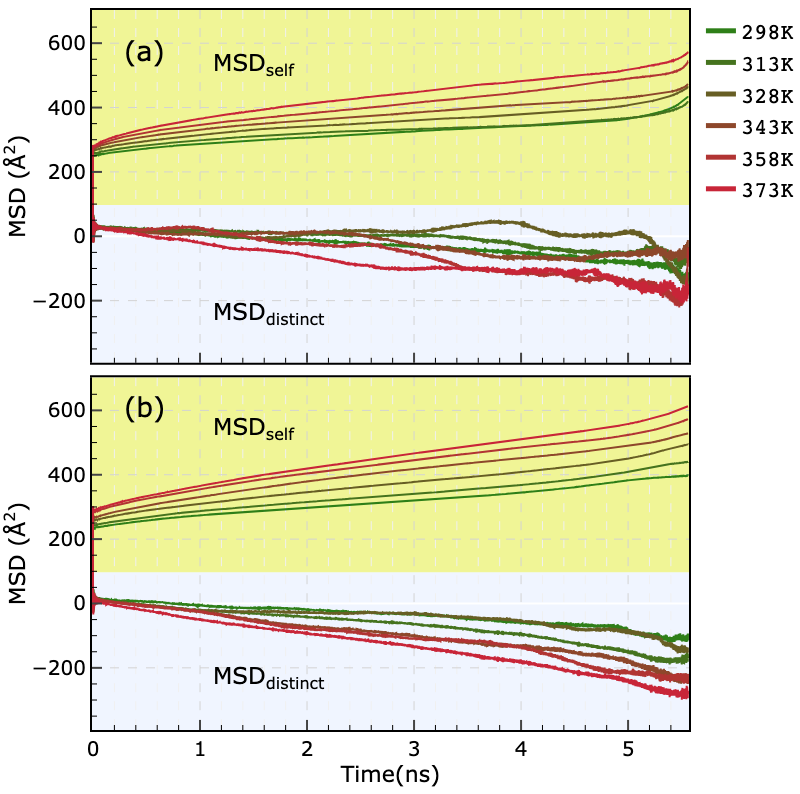}
		\caption{Time-evolution of mean squared displacement (MSD) of Li atoms calculated at  298, 313, 328, 333, 358, and 373K for uncorrelated, $\mathrm{ MSD}_{\mathrm{self}} $, and correlated part, $\mathrm{ MSD}_{\mathrm{distinct}} $. (a) and (b) are MSDs calculated from DLMD simulation with U=0.0 eV and U = 4.0 eV, respectively. }
		\label{fig:Figure4}
	\end{figure}
	
	\begin{figure*}[htbp]
		\centering
		\includegraphics[width=0.9\textwidth]{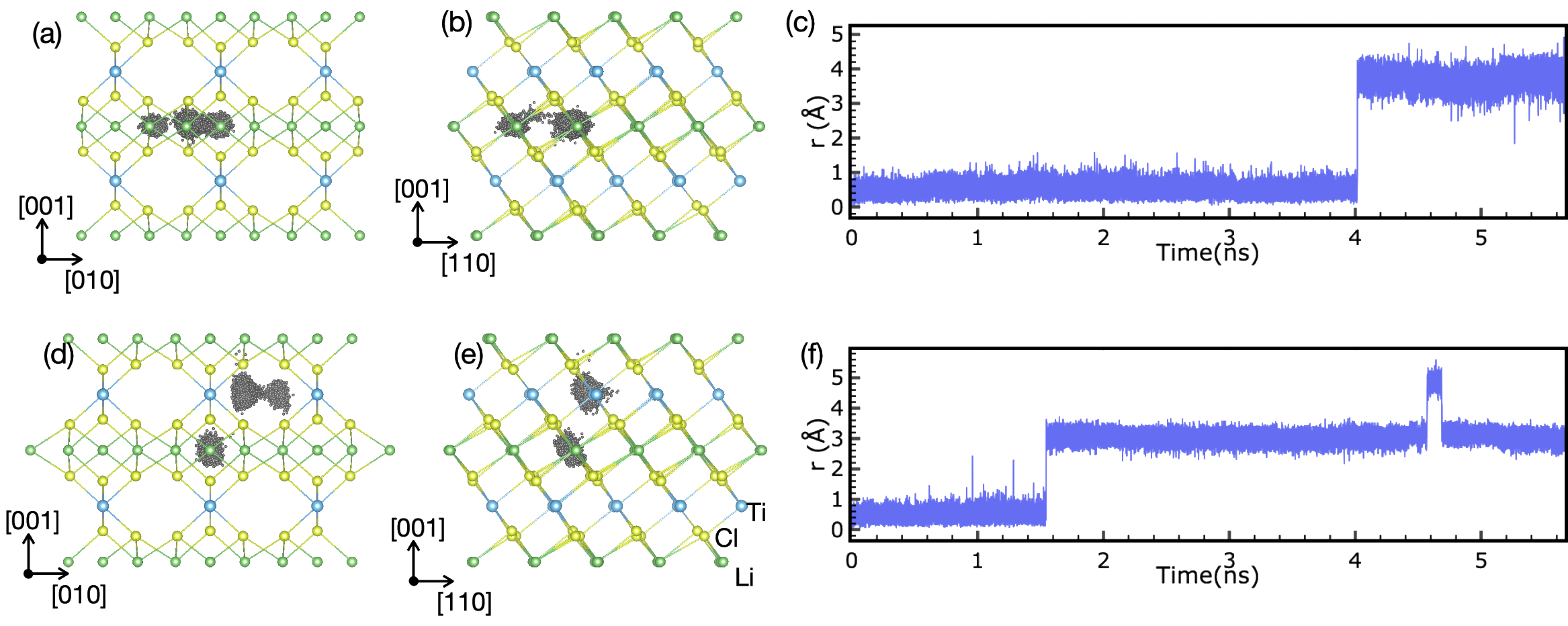}
		\caption{Illustration of intra-layer Li-ion movement (gray balls) along Li$_3$TiCl$_6$  (a) [010], and (b) [110] view-point directions. (c) Time-evolution of Li atom translations corresponding to (a) and (b) as a function of time. Illustration of inter-layer Li-ion movement (gray balls) along Li$_3$TiCl$_6$  (d) [010], and (e) [110] view-point directions. (f) Translation distance of Li atoms corresponding to (d) and (e) as a function of time.}
		\label{fig:Figure5}
	\end{figure*}
		\subsection{Diffusion coefficients}
	The diffusion of Li-ions was determined from DLMD trajectories, where the positions of Li atoms, denoted as $\left\{\boldsymbol{r_i(t)}\right\}_{i=1}^{N} $ , are tracked as a function of time $t$ for all $N$ Li atoms. This involves calculating the mean square displacement of Li atoms, $\mathrm{ MSD}_{\mathrm{Total}} $ \cite{msd, anti_msd}  as follows.
	\begin{equation}
		%\begin{split}
		\mathrm{MSD}_{\mathrm{total}} = \frac{1}{N} \Bigg|\sum_{ i=1}^{N} \Delta \boldsymbol{r}_i(t)\Bigg| ^2 
		%\end{split}
	\end{equation}
	We compute uncorrelated motion of MSD as follows. 
	\begin{equation}
		%\begin{split}
		\mathrm{MSD}_{\mathrm{self}} = \frac{1}{N} \sum_{ i=1}^{N} |\Delta \boldsymbol{r}_i(t)| ^2 
		%\end{split}
	\end{equation}
	From Eq. (6) and Eq. (7), distinct part of MSD is calculated by
	\begin{equation}
		%\begin{split}
		\mathrm{MSD}_{\mathrm{distinct}} = \mathrm{MSD}_{\mathrm{total}}  - \mathrm{MSD}_{\mathrm{self}} 
		%\end{split}
	\end{equation}
	
	When $\mathrm{ MSD}_{\mathrm{distinct}}$ $<$ 0 describes negative correlation between Li-ions and, hence, negatively affect the Li-ion transport. Based on Equations (6-8), we calculated $\mathrm{ MSD}_{\mathrm{self}} $ and $\mathrm{ MSD}_{\mathrm{distinct}} $ and collect the results in FIG.\ref{fig:Figure4}. It is seen that $\mathrm{ MSD}_{\mathrm{distinct}}$  values for Li$_3$TiCl$_6$  at different temperatures show negative correlation (or anti-correlation) between Li atoms during their dynamics. To quantify this observation the diffusion coefficient was calculated on the basis of MSD using
	
	\begin{equation}
		D=\frac{{\rm{MSD}}(\Delta t)}{2d\Delta t}
	\end{equation}
	where $d$ is dimensionless quantity which is equal to 3 for three dimensional transport. Since,  Li$_3$TiCl$_6$ has anti-correlated Li-ion dynamics, calculated $D_{self}$ is highly overestimated about 14 times of $D_{Total}$, which is $1.6 \times 10^{-14} cm^2/s$ at 298K with DLP generated from DFT+U = 4.00 eV. 
	
	One way to understand this phenomenon is by analyzing Haven’s ratio (H$_R$) \cite{haven_ratio} which is defined as the ratio of the self-diffusion coefficient of Li-ions ($\mathrm{ D}_{\mathrm{self}}$) to the total diffusion coefficient ($\mathrm{ D}_{\mathrm{Total}}$) using Eq. 9. 
	\begin{equation}
		H_R =  \frac{\mathrm{ D}_{\mathrm{self}}}{\mathrm{ D}_{\mathrm{Total}}}
	\end{equation}
	Here, $D_{self}$ represents the individual motion of ions which is independent of overall charge transport. On the other hand, $D_{Total}$ is the diffusion coefficient calculated from the material's conductivity by considering the collective motion of ions that contributes to overall charge transport. In the case of uncorrelated Li-ion dynamics, $\mathrm{ D}_{\mathrm{Total}} \equiv \mathrm{ D}_{\mathrm{self}}$, and $H_R$ becomes 1. If $H_R$ is less than 1, $\mathrm{ D}_{\mathrm{distinct}}$ should have been positively correlated with Li-ion dynamics, typically observed in liquid and glass electrolytes.\cite{anti_msd, haven_ratio} In our case, since $H_R$ is greater than one, we propose that Li-ion dynamics may involve an interstitial and/or inter-layer transport mechanism.
	
	%\begin{figure}[htbp]
	%	\centering
	%	\includegraphics[width=0.45\textwidth]{./energy_barrier.png}
	%	\caption{Energy barrier in eV }
	%	\label{fig:your_label}
	%\end{figure}
	\subsection{Transport mechanism}
	Furthermore, we analyzed the Li-ion path to confirm the inter- and intra-layer diffusion of the ions. Among all the Li atoms in the trajectory simulated at 298 K, we arbitrarily selected two Li atoms to track the ionic motion over time (in FIG.\ref{fig:Figure5}). Figures 5a and 5b clearly illustrate the Li atom (gray-colored dots) moving within the Li layer through interstitial hopping. Simultaneously, inter-layer motion between Li- and Ti-layers is also observed through interstitial migration, as shown in Fig. 5d and 5e.
	
	The movements of ionic movement corresponding to intra- and inter-layer migration are shown in Fig. 5c and 5f. More interestingly, Fig. 5f demonstrates that inter-layer Li-ion migration reaches the third cation layer at 4.7 ns through interstitial sites, proving multiple-site hopping of Li-ion. Since the Li$_3$TiCl$_6$ crystal structure has partially occupied Ti-2$a$ as well as Ti-4$g$, the structure possesses inherent voids that can also possibly act as hopping sites for Li-ion migration. Therefore, our study confirms that Li$_3$TiCl$_6$ has an inter- and intra-layer interstitial hopping-based Li-ion transport mechanism.

	\begin{figure}[htbp]
		\centering
		\includegraphics[width=0.47\textwidth]{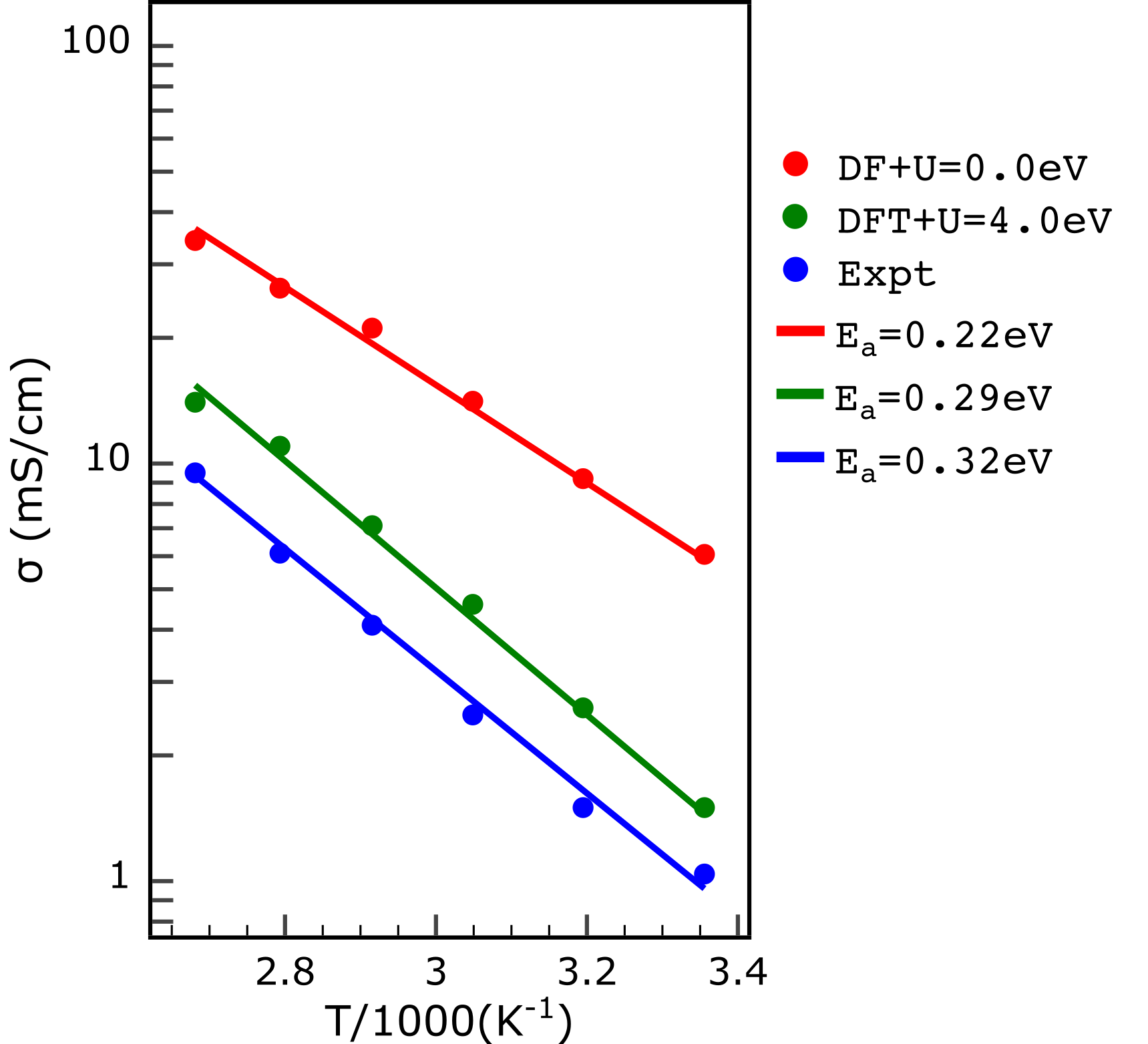}
		\vspace{2mm}
		\caption{Arrhenius plot for ionic conductivity as a function of temperature calculated using of DLMD at 298, 313, 328, 333, 358, and 375 K with U = 0.0 and U = 4.0 eV in comparison with experimental data \cite{wang2023li3ticl6}. }
		\label{fig:Figure6}
	\end{figure}
	\subsection{Ionic conductivity}
	The ionic conductivity, denoted as $\sigma$, can be determined from the self-diffusion coefficient using the Nernst–Einstein equation\cite{sigma}:
	\begin{equation}
	\sigma = \frac{ne^2Z^2}{H_Rk_{\mathrm{B}}T}\mathrm{ D}_{\mathrm{self}} 
\end{equation}
	Here, $n$ represents the ion density of Li, $e$ is the elementary electron charge, $Z$ denotes the valence of Li, and $k_{\mathrm{B}}$ is the Boltzmann constant. The calculated ionic conductivity values from diffusion coefficients align well with experimental values \cite{wang2023li3ticl6}, as demonstrated in FIG.\ref{fig:Figure6}. Additionally, when considering the temperature dependence of ionic conductivity in solid-state electrolytes, high-temperature ionic conductivities obtained from DLMD simulations can be leveraged to estimate the activation barrier of the electrolytes at lower temperatures using the Arrhenius relationship: $\sigma = \sigma_0e^{-E_{\mathrm{a}}/(k_{\mathrm{B}}T)}$. The calculated activation energy is 0.29 eV, closely matching the experimental value of 0.32 eV. In addition, Li-ion migration barriers were calculated using the DFT-based NEB method along different crystallographic directions. The barriers along [110], [101], [010], [100] are 0.30, 0.303, 0.312, and 1.14 eV, respectively. The inter-layer ([110] and [101]) migration barriers are considered along interstitial sites and are very close to the activation energy calculated in DLMD simulations. Thus, the calculated activation energy and Li-ion barrier energy confirm the accuracy of both DLMD and DFT simulations, respectively.
	
	\section{Conclusion}
	We performed large-scale and MLFF-driven molecular dynamics simulations to investigate the Li-ion transport mechanism in cation-disordered Li$_3$TiCl$_6$ cathode at six different temperatures, ranging from 298K to 373K. Deep neural network method along with data generated by  AIMD simulation were used to build a high-fidelity MLFF. Predicting accuracy of atomic forces, energy, and structure by our trained MLFF was confirmed with set of new AIMD data  and corresponding RDF. The calculated self and distinct part of Li-ion MSD reveal that the Li-ions are involved in anti-correlation motion that was rarely reported for solid-state materials. 
	
	In the same way, analysis of trajectory from DLMD infers that the Li-ion transportation occurs through interstitial hopping which was confirmed by intra- and inter-layer Li-ion movement as a function of simulation time. The temperature dependent  ionic conductivity and, thus, activation barrier values for Li$_3$TiCl$_6$ demonstrate a decreasing trend with temperature, aligning with typical behavior of ionic conductors. Moreover, activation energy of 0.29 eV, which is in close agreement with experimental result, matches well with inter-layer ionic diffusion barrier calculated by DFT along [110] crystallographic direction. Overall, the combination of machine-learning methods and AIMD simulations elucidates the complex Li-ion electrochemical properties of the Li$_3$TiCl$_6$ cathode by significantly reducing computational time. Hence, our work strongly suggests that the MLFF using deep neural networks could be promising for studying large-scale complex materials.
	
	\section*{Acknowledgments}
	This work was supported by the by the Assistant Secretary for Energy Efficiency and Renewable Energy, Office of Vehicle Technologies of the US Department of Energy, through the Battery Materials Research (BMR) program. We gratefully acknowledge the computing resources provided on Bebop, a high-performance computing cluster operated by the Laboratory Computing Resource Center at Argonne National Laboratory.
	
	%\section*{Author Contributions}
	%A.B. and C.D. conceived the idea and designed the study. A.B. performed the deep learning molecular dynamics simulations and analyzed the data. C.D. conducted the density functional theory calculations. A.B. and C.D. co-wrote the manuscript.
	
	\section*{Conflict of Interest}
	The authors declare no conflict of interest.

	%%%%%%%%%%%% Supplementary Methods %%%%%%%%%%%%
	%\footnotesize
	%\section*{Methods}
	
	%%%%%%%%%%%%% Acknowledgements %%%%%%%%%%%%%
	%\footnotesize
	%\section*{Acknowledgements}
	
	%%%%%%%%%%%%%%   Bibliography   %%%%%%%%%%%%%%
	\normalsize
	\bibliography{references}
	
	%%%%%%%%%%%%  Supplementary Figures  %%%%%%%%%%%%
	%\clearpage
	
	%%%%%%%%%%%%%%%%   End   %%%%%%%%%%%%%%%%
	%\end{multicols}  % Method B for two-column formatting (doesn't play well with line numbers), comment out if using method A
\end{document}